# Diagnosis of rotator cuff lesions by FT-Raman spectroscopy: a biochemical study

Sergio Godoy Penteado[1], Cláudio S. Meneses[2], Anderson de Oliveira Lobo[1], Airton Abrahão Martin[1], *Herculano da Silva Martinho[1]

[1] Laboratory of Biomedical Vibrational Spectroscopy,
Institute of Research and Development - IP&D,
[1] University of the Valley of Paraíba - UniVap, Av. Shishima Hifumi, 2911, 12244-000

[2] CIPAX – Medical Diagnosis
Av. Nove de Julho, 507, 12243-000

São José dos Campos, São Paulo, Brazil

E-mail: godoypenteado@terra.com.br, claudio@cipax.com.br, anderson@univap.br, amartin@univap.br, *hmartinho@univap.br

**Abstract** The biochemical changes on normal and degenerated tissues of rotator cuff supraspinatus tendons were probed by FT-Raman spectroscopy. The Raman spectra showed differences on the spectral regions of cysteine, amino acids, nucleic acids, carbohydrates, and lipids. These spectral differences were assigned to pathological biochemical alterations due to the degenerative process of the tendon. Principal Components Analysis was performed on the spectral data and enabled the correct classification of the spectra as normal (grade 1) and degenerated (grades 2 and 3). These findings indicate that Raman spectroscopy could be a very promising tool for the rotator cuff supraspinatus tendon diagnosis and for quantification of their degenerative degree.

**Keywords** Raman spectroscopy, rotator cuff, collagen, tendon degeneration, cysteine

**Introduction**

The rotator cuff is a group of four muscles (supraspinatus, infraspinatus, teres minorand, and subscapularis) that in conjunction with their tendons act stabilizing the glenohumeral joint during shoulder movement.[1,1,2,2]. The rotator cuff pathologies could vary as tendinosis, tendonitis, partial tear or complete tear. The supraspinatus tendon localized at the superior part of the cuff is the most directly affected by these tendinopathies.[2]

The main etiologic cause of the rotator cuff pathologies include cuff impingement,[3] degenerative changes,[2] overuse and trauma. [4] All these factors are influenced by age and personal lifestyle of the patient, as well as predisposal. However, they are poorly understood both from the cellular and molecular perspective.[2] It is consensus that the overuse and trauma could accelerate or start the degenerative process.

Healthy tendons are constituted mainly by type I collagen (65% - 80%).[2,5] Types III and V collagens are also present but in low amounts. The others components are water, elastic fibers and proteoglycans. [2,5]. Morphologically the healthy tendons are white brilliant and have a fibroelastic texture while the degenerated tendons lose their white appearance becoming grey-brown and amorphous.[2] The collagen degeneration in tendons is known as *tendinosis* while *tendinitis* refers to the inflammatory process of tendons. In spite of their general use, these terms could be correctly assigned to a specific tendon, only after a histopathological analysis.[2].

The conventional diagnosis of the rotator cuff injuries is based on morphological pieces of information given by arthrography, radiograph, magnetic resonance imaging and ultrasound.[4] The double contrast arthrography had reports of 100% diagnosis accuracy. [4] The others techniques have sensibility and specificity limitations depending on the degeneration stage and/or pathology etiology. However, none of these techniques gives biochemical details what enables one to establish either the degree of the tendon degeneration or the tendon quality.

Raman spectroscopy is an optical technique that provides information about the molecular vibrational degrees of freedom of the investigated sample being widely used for quantitative and qualitative analytical studies in the fields of chemistry, geology, pharmacology, and condensed matter physics. Recently, it has emerged as a nondestructive analytical tool for the biochemical characterization of biological systems due to several advantages as sensitivity to small structural changes, non-invasive sample capability, minimal sample preparation, and high spatial resolution in the case of Raman microscopy. This technique does not require wide sample preparation or pre-treatment, and making use of the FT-Raman technique employing light sources in the infrared region, the detection of weak Raman signals become easier due to fluorescence suppression. Moreover, the excitation in the near infrared, at 1064 nm, also minimizes the photo-degradation of the sample, allowing the employment of larger power densities to compensate the weak Raman signal generated by longer wavelengths.[6]

The goal of this work is show that Raman spectroscopy can be used as a very effective tool for the rotator cuff supraspinatus tendon diagnosis. Due to it biochemical sensibility it could provide specific information concerning the grade of degeneration of the tendon and be complementary to other invasive techniques, e.g., arthroscopy, enabling morpho-biochemical diagnosis. For the patients it is relevant since the quantification of the tendon quality make possible better prognostic implying better life quality. It is important to point out that from the best of our knowledge, similar study is absent in the literature.

**Methodology**

    **A. Tissues**

This research was done following ethical principles established by the Brazilian Federal Healthy Ministry. The patients were informed about the subject of the research and signed permission for collecting their tissue samples.

The rotator cuff supraspinatus tendon tissues samples were obtained from 7 patients submitted to shoulder surgery at the Clinica "São Gabriel", São José dos Campos, São Paulo, Brazil. The samples, soon after the surgical procedure, were identified, snap frozen and stored in liquid nitrogen (77 K) in cryogenic vials (Nalgene ®) before the FT-Raman spectra recording. Each tissue was sectioned into 2 or 3 partitions depending on their size totalizing 19 samples. The Raman spectra were measured on 3 different points on each sample and those taken near to the distal or ruptured tendon were discharged (7 spectra). Soon after, the samples were fixed in 10 % formaldehyde solution, to further histopathological analysis.

Each measured sample was histopathologically assessed and classified according to the tendon degenerative grading scale proposed by G. P. Riley *et al*.[7] This scale has 4 graduations:

- *Grade 1* (normal tendon): It is characterized by a wavy outline of collagen fibers, individual fibers easily discernible, tenocyte nuclei elongated, and parallel to bundles of collagen.

- *Grade 2* (mild degeneration): It is characterized by collagen fibers relatively well aligned but with irregular waviness. Individual fibers not readily identified, tenocyte cell nuclei shorter and nuclei in Indian file.

- *Grade 3* (moderate degeneration): At this grade begins a process of collagen hyalinization, loss of orientation of the fibers and tenocyte nuclei, and increase of the number of cell nuclei.
- *Grade 4* (severe degeneration): The collagen fibers have a diffuse hyalinization of homogeneous appearance. The tenocyte has rounded cells, complete loss of orientation of the fibers and the number of nuclei reduced.

### B. FT- Raman spectroscopy

A FT-Raman spectrometer (Bruker RFS 100/S) was used with an Nd:YAG laser at 1064 nm as excitation light source. The laser power at sample was kept 110 mW while the spectrometer resolution was set to 4 cm$^{-1}$. The spectra were recorded with 300 scans.

For FT-Raman data collection, samples were brought to room temperature and kept moistened in 0.9 % physiological solution to preserve their structural characteristics, and placed in a windowless aluminum holder for the Raman spectra collection. We notice that the chemical species present in the physiological solution ($Ca^{2+}$, $Na^+$, $K^+$, $Cl^-$, water) do not have measurable Raman signal and their presence do not affect the spectral signal of the tissues.

### C. Data analysis

All spectra were baseline corrected and normalized to the 1445 cm$^{-1}$ band intensity. This band corresponds to C—H deformation mode of methylene group and it is nearly conformational insensitive. For this reason, it is a good standard for biological Raman spectral normalization. After that was taken the mean spectra for each sample (typically over 3 spectra). The spectral differences were statistically analyzed by Principal Components Analysis (PCA). The PCA analysis was performed on the statistical software

Minitab 14.20 (Minitab Inc.) over the range 300 – 1800 cm$^{-1}$ by computing the covariance matrix.

**Results and Discussion**

Figure 1 shows some representative normalized mean Raman spectra of rotator cuff tendons with grades 1, 2 and 3 of degeneration. No severe (grade 4) degeneration was observed in the analyzed patients. The dashed curves correspond to spectra from samples with cartilaginous synovium contamination as pointed out by the histopathological analysis. In the clinical routine the signal of synovium will be always present and for this reason their Raman spectra could not be discharged. The spectra were vertically translated for clarity. In the Table 1 are presented the major vibrational assignments for the Raman bands observed on tendon tissues.[8,9]

The relative spectral differences, $\frac{S_{path} - S_n}{S_n}$, between pathological and normal tendons were calculated to get a clear view of the main spectral differences between pathological and normal samples. $S_{path}$ and $S_n$ are the Raman spectra of pathological and normal tissues, respectively. These differences are shown in the Figures 2 and 3.

Figure 2 shows the spectral differences between 200 and 800 cm$^{-1}$. It is clear that the collagen-related bands (amino acids and proteins) dominate the spectral variation. The major spectral difference in relation to the normal spectra was found in the 200-400 cm$^{-1}$ region which corresponds to the cysteine vibrational bands. This amino acid is present in the type III collagen. Thus this finding is coherent with the pathology description since it is well known that degenerated tendons have increased amounts of type III collagen.[5,7]

Figure 3 shows the spectral differences between 850 and 1800 cm$^{-1}$. In this region the spectral differences were due to proline, valine, phenylalanine, lipids, carbohydrates, nucleic acids and collagens alterations. The amino acids content changes are probably related to the collagen and elastin degenerations. The alterations in the number of tenocyte cell nuclei in the tendinopathies [5,7] are probably responsible for the nucleic acids bands variations. Changes on the amount of lipids and carbohydrates are also commonly related to tendon degeneration.[2]

In order to establish whether these spectral alterations enable the tissue classification as normal and degenerate tendons, all FT-Raman spectra were analyzed through the PCA technique. Table 2 shows the eingenvalues and contributions of the first 10 PCA factors. The first 6 factors are responsible for 100% of spectral variability.

Figure 4 shows the loading plots for the first 4 factors. These factors are responsible for 99,5% of spectral variability. The set of peaks of the Fig. 4a (first score) resembles the set of bands seen in the normal tendon (Fig. 1). On the other side many peaks seen on the Fig. 4b (second score) have opposite concavity and shifted position as compared to the first score, indicating that this PC component is sensible to the spectral differences originated on the pathologic tissues. This fact could also be verified when one compares the peaks of Fig. 4b to the spectral difference plotted on Fig. 2 and 3. The plots of Fig. 4c and 4d (third and forth scores) present a lot of peaks that not have correspondence to the raw spectral data. These peaks correspond to the fine structure of the raw-data being related to the experimental noise. Thus, it is possible to argue that the main spectral differences seen in the data could be recognized by retained only the first and second PC components.

Figure 5 shows the Second against First PCA components plot. This plot shows a clear discrimination between normal (grade 1) and lesion tendons (grades 2 and 3). The dot line is a guide for the eyes that shows the limit of the normality region. It is noticed that had not been found clear discrimination between groups with grade 2 and 3 degeneration and only one false negative was observed. The calculated sensibility for this method was 95 % while the specificity was 100 % indicating that the Raman spectroscopy could be a very promising tool for tendon lesions diagnosis.

**Conclusions**

The preliminary results presented on this work indicate that the Raman spectroscopy is a sensible technique to the main biochemical alterations occurring on the degenerative process of the rotator cuff supraspinatus tendon. The main biochemical changes were observed on the cysteine, amino acids and nucleic acids bands. The cysteine is present on type III collagen while amino acids are the constituents of collagens in general. The nucleic acids are present on the nuclei of tenocyte cell. These alterations are compatible with the pathology alterations reported on the literature. The reports indicate that there is an increasing of type III collagen amount following the degenerative process, as well as, alteration in the nuclei cells number. Alterations on lipids and carbohydrates are also expected to occur and were also observed on the spectroscopy data. The PCA analysis performed indicates that the biochemical alterations probed by Raman were sensible enough to classify normal (grade 1) and degenerated (grades 2 and 3) tendons what indicated the viability of applying this technique to tendon diagnosis. To enforce the diagnosis possibility more experimental efforts are needed in order to increase the spectrum data base and perform more accurate statistical methods in order to discriminate all grades of degeneration.

**Acknowledgements**

A.A.M. thanks CNPq (302393/2003-0) and FAPESP (2001/14384-8) for providing financial support.

TABLES

Table 1. Vibrational assignment of the main spectral regions observed in normal and pathological tendon tissues.

| Band (cm$^{-1}$) | Vibrational Mode | Major Assignment |
|---|---|---|
| 218 | CCS$_{bend}$ | cysteine – collagen (III) |
| 250 | CCC$_{rock}$ | cysteine– collagen (III) |
| 305 | CCC$_{bend}$ | cysteine– collagen (III) |
| 345 | SH$_{tors}$ | cysteine– collagen (III) |
| 360-500 | NH$_{3\ tors}$, CCN$_{bend}$ | amino acids |
| 540 | SS bridges | cysteine |
| 565 | CO$_{2\ rock}$ | amino acids |
| 590-630 | CC$_{str}$ | amino acids |
| 725 | CS$_{str}$ | cysteine |
| 740-800 | CH$_{2\ rock}$, CO$_{2\ bend}$ | amino acids |
| 816 | CO$_{2\ wagg}$ | amino acids |
| 850-900 | CC$_{str}$, ring breathing, OPO$_{str}$ | proline, tyrosine, DNA |
| 920-975 | CC$_{st}$, α-helix | proline, valine, protein conformation, glycogen |
| 1005 | symmetric ring breathing mode | phenylalaline |
| 1015-1150 | CC$_{str}$, CO$_{str}$, PO$_{2\ str}$, CN$_{str}$, OPO$_{str}$ | lipids, nucleic acids, proteins, carbohydrates |
| 1250 | CN$_{str}$ of amide III | polar triple helix of collagen, elastin |
| 1270 | CN$_{str}$ of amide III | non-polar triple helix of collagen, elastin |
| 1320 | CH$_3$,CH$_{2\ twist}$ | collagen (I) |
| 1347 | CH$_3$,CH$_{2\ wagg}$ | collagen (I) |
| ~ 1450 | CH$_{2\ twis}$ | lipids, carbohydrates, proteins and pentose |
| ~1600 | CO$_{2\ str.\ asym.}$ | amino acids |
| ~1640 | amide I | collagen |
| ~1670 | amide I | collagen, elastin |

Table 2. The eigenvalues and % contribution of the first 10 PCA factors analysis.

| Factor number | Eigenvalue | % Contribution |
|---|---|---|
| 1 | 297,82 | 97,4 |
| 2 | 3,80 | 1,2 |
| 3 | 1,89 | 0,6 |
| 4 | 0,90 | 0,3 |
| 5 | 0,61 | 0,2 |
| 6 | 0,38 | 0,1 |
| 7 | 0,07 | 0 |
| 8 | 0,04 | 0 |
| 9 | 0,03 | 0 |
| 10 | 0,02 | 0 |

FIGURE CAPTIONS

Figure 1. Representative mean Raman spectra of supraspinatus rotator cuff tendons with grades 1, 2 and 3 of degeneration.

Figure 2. Relative spectral difference of representative spectra between 200 and 800 cm$^{-1}$.

Figure 3. Relative spectral difference of representative spectra between 850 and 1800 cm$^{-1}$.

Figure 4. Loading plots for first (a), second (b), third (c) and fourth (d) scores.

Figure 5. Second vs First PCA components of normal and lesion tendons.

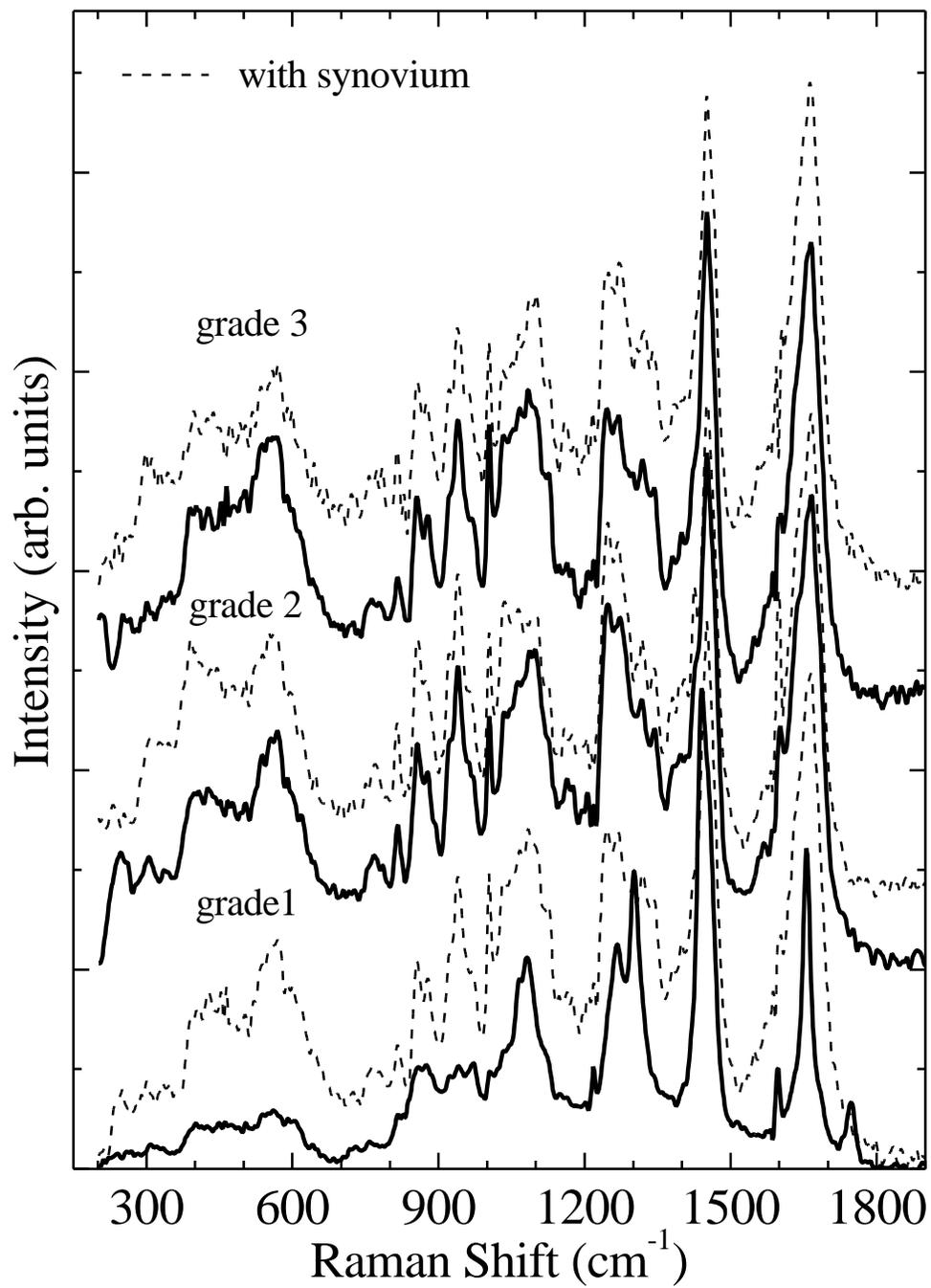

Figure 1

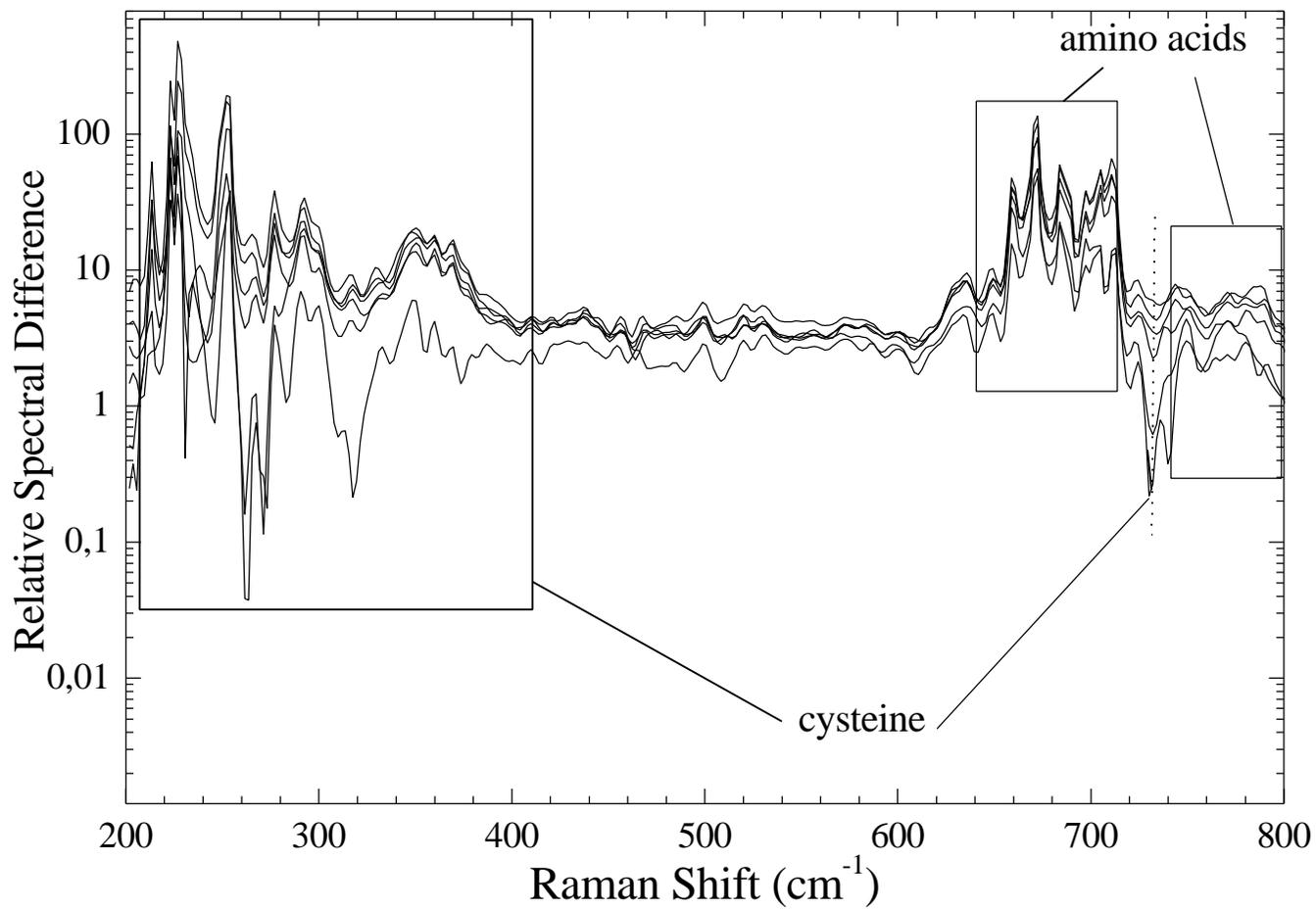

**Figure 2**

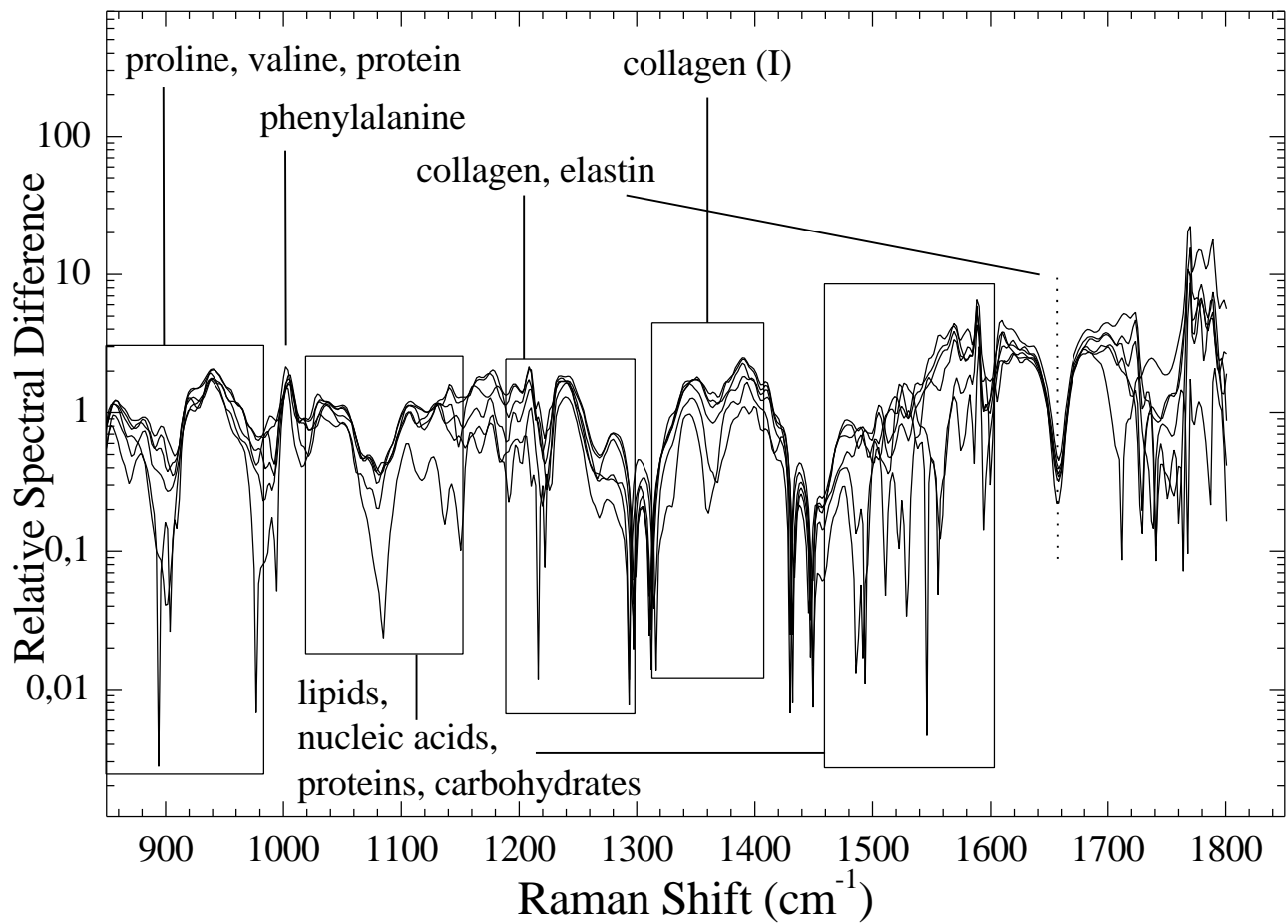

**Figure 3**

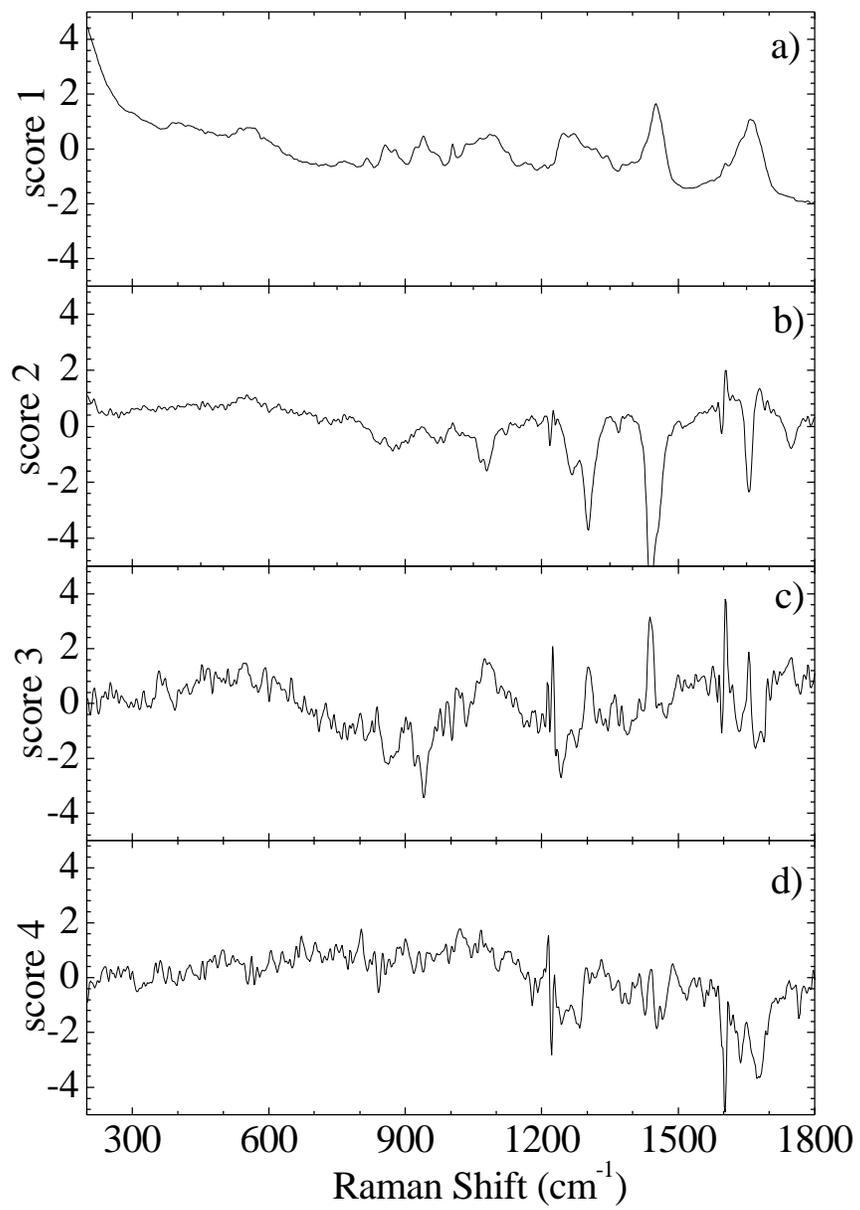

Figure 4

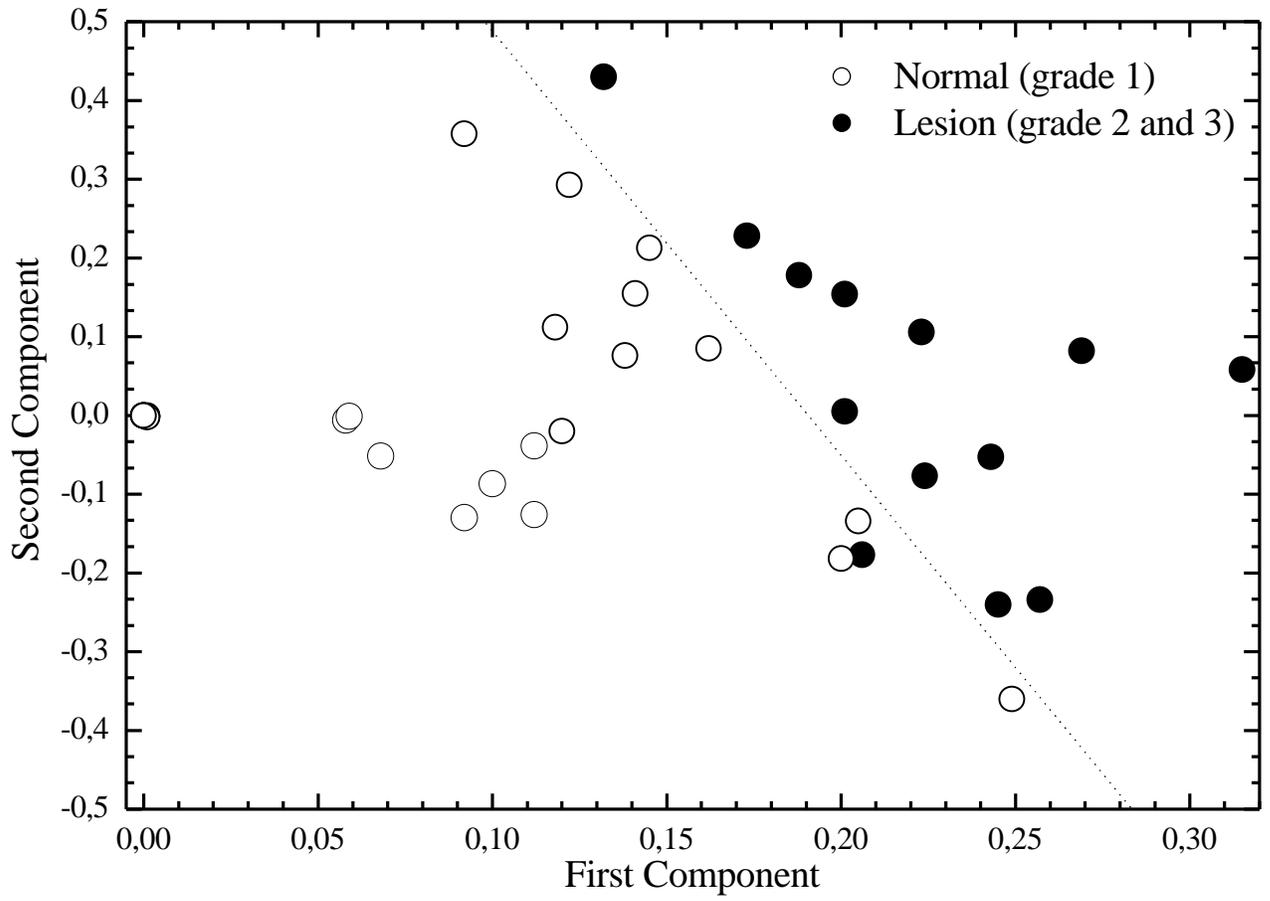

Figure 5